\documentclass[aip,jap,reprint]{revtex4-1}

\usepackage{graphics}
\begin{document}

\title{Effect of local environment on crossluminescence kinetics in SrF$_2$:Ba and
CaF$_2$:Ba solid solutions}

\author{M.A. Terekhin}
\affiliation{P.N. Lebedev Physical Institute, Leninskij Prospekt 53,
119991 Moscow, Russia}

\author{V.N. Makhov}
\email[]{makhov@sci.lebedev.ru}
\affiliation{P.N. Lebedev Physical Institute, Leninskij Prospekt 53,
119991 Moscow, Russia}

\author{A.I. Lebedev}
\affiliation{M.V. Lomonosov Moscow State University, Moscow 119991, Russia}

\author{I.A. Sluchinskaya}
\affiliation{M.V. Lomonosov Moscow State University, Moscow 119991, Russia}


\begin{abstract}
Spectral and kinetic properties of extrinsic crossluminescence (CL) in
SrF$_2$:Ba(1\%) and CaF$_2$:Ba(1\%) are compared with those of intrinsic CL in
BaF$_2$ and are analyzed taking into account EXAFS data obtained at the Ba
$L_{\rm III}$ edge and results of first-principles calculations. The CL decay
time was revealed to be longer in SrF$_2$:Ba and CaF$_2$:Ba compared to BaF$_2$.
This fact contradicts the expected acceleration of luminescence decay which
could result from an increased overlap of wave functions in solid solutions due
to shortening of the Ba--F distance obtained in both EXAFS measurements and
first-principles calculations. This discrepancy is explained by the effect of
migration and subsequent non-radiative decay of the Ba\,($5p$) core holes in
BaF$_2$ and by decreasing of the probability of optical transitions between
Ba\,($5p$) states and the valence band in SrF$_2$:Ba and CaF$_2$:Ba predicted
by first-principles calculations.

\texttt{DOI: 10.1016/j.jlumin.2015.05.022}
\end{abstract}


\maketitle 

\section{Introduction}

It is well known that so-called crossluminescence (CL),~\cite{1,2} also known
as core-valence luminescence~\cite{3} or Auger-free luminescence,~\cite{4} is
due to radiative recombination of electrons from the valence band with the holes
created in the uppermost core band. CL is observed in a number of wide band-gap
ionic crystals in which the Auger decay of these holes is energy forbidden. For
BaF$_2$ crystal, an example of the well-known CL-active material, CL can be
represented as radiative electronic transitions from the F$^-$\,($2p$) valence
band to the Ba$^{2+}$\,($5p$) core band; although it is generally accepted that
the core hole becomes self-trapped before the radiative CL transition takes
place. In BaF$_2$ such transitions give a rather bright broadband luminescence
in the ultraviolet region with a maximum at 220~nm.

The important characteristic of CL is its short decay time $\tau \sim 1$~ns. This
property of CL has resulted in a rather strong interest to CL crystals as possible
fast scintillators in previous years although up to now CL-based scintillation
detectors have found limited applications (see, e.g., Refs.~\onlinecite{5,6}).
However, nowadays the interest to CL scintillators has reappeared because of
increasing demand for ultrafast radiation detectors for high-energy physics
experiments at high-luminosity accelerators and for positron emission tomography
based on time-of-flight technique.~\cite{7}

Although CL was discovered more than 30 years ago, many features of CL mechanism
are still not well understood. In particular, the real lattice configuration of
the initial state for CL transition is under discussion (see, e.g.,
Ref.~\onlinecite{8}). In order to get some insights into this problem, in the
present work the influence of local environment around the core-hole on the
excitation spectra and kinetics of CL has been studied for so-called extrinsic
(impurity) CL. If some alkali or alkaline earth halide crystal (e.g. SrF$_2$ or
CaF$_2$) which does not have intrinsic CL is doped with a heavier metal ion
(e.g. Ba$^{2+}$), then the radiative recombination of the valence electrons with
holes created in the uppermost $np$ levels of the impurity ion ($5p$ levels of
Ba$^{2+}$) can become possible.~\cite{9}  Of course, such extrinsic CL is
observed if the spectrum of these transitions is in the optical transmission
region of the host crystal. The excitation threshold of the impurity CL coincides
with the energy separation between the impurity $np$-level and the bottom of
the conduction band.

It is well known that crystals with the fluorite structure (CaF$_2$, SrF$_2$,
and BaF$_2$) undergo a phase transition from cubic fluorite to orthorhombic
PbCl$_2$ structure at high pressure.~\cite{10,11,12}  Thus, if Ba$^{2+}$ ions are
incorporated into the SrF$_2$ or CaF$_2$ host having smaller lattice parameter
than BaF$_2$, the strong local stress around the Ba$^{2+}$ ions can result in the
appearance of a local phase transition. Accordingly, the local environment of
the Ba$^{2+}$ ions in these hosts can become very different from that in BaF$_2$
and can remarkably modify the properties of Ba$^{2+}$-related impurity CL in
SrF$_2$ and CaF$_2$ compared to those of intrinsic CL in BaF$_2$. The
conventional experimental technique for studying local environment of atoms
in solids is extended X-ray absorption fine-structure (EXAFS) spectroscopy
which provides, in particular, data on interatomic distances and coordination
numbers, i.e. can directly indicate whether the local phase transition takes
place.

In the present work the local structure around the Ba$^{2+}$ ions in SrF$_2$:Ba(1\%)
and CaF$_2$:Ba(1\%) crystals has been studied using EXAFS technique. The
obtained data were compared with the results of first-principles calculations
of the local structure distortions around the Ba$^{2+}$ ions in SrF$_2$:Ba and
CaF$_2$:Ba solid solutions. The excitation spectra and kinetics of extrinsic
CL in SrF$_2$:Ba and CaF$_2$:Ba were measured using synchrotron radiation.
Finally, the observed spectral and kinetic properties of extrinsic CL in
SrF$_2$:Ba and CaF$_2$:Ba are analyzed taking into account the results of
EXAFS measurements and first-principles calculations.

\section{Experimental details}

The EXAFS measurements were carried out on station 7.1~\cite{13} of synchrotron
radiation source (SRS) at Daresbury Laboratory with an electron beam energy of
2~GeV and the maximum storage ring current of 250~mA in multi-bunch mode. The
incident radiation was monochromatised using a double-crystal Si(111)
monochromator. Experiments were performed at 300~K at the Ba $L_{\rm III}$
edge (5.247~keV) in the energy range 5.15--5.62~keV (the upper boundary was
limited by the proximity of the Ba $L_{\rm II}$ edge). The EXAFS data were
collected in transmission mode using two ion chambers filled with He+Ar gas
mixtures that gave 20\% and 80\% absorption of incident radiation.

The oscillatory EXAFS function $\chi(k)$ was extracted from the absorption curve
$\mu x(E) = \ln(I_0/I)$ (where $E$ is the energy of radiation) in the normal
way.~\cite{14}  After removing the pre-edge background, splines were used to
extract the smooth atomic part of the absorption $\mu x_0(E)$, and the dependence
$\chi = (\mu x - \mu x_0)/\mu x_0$ was calculated as a function of the
photoelectron wave vector $k = [2m (E - E_0)/\hbar^2]^{1/2}$. The energy origin
$E_0$ was taken to be the position of the inflection point on the absorption
edge. The edge steps ranged from 0.012 to 0.019. For each sample direct and
inverse Fourier transforms with modified Hanning windows were used to extract
the information about the first three shells from the experimental curve
$\chi(k)$. The distances $R_j$, coordination numbers $N_j$, and Debye-Waller
factors $\sigma_j^2$ for each shell as well as the energy origin correction
$\delta E_0$ were simultaneously varied to obtain the minimum root-mean-square
deviation between the experimental and calculated $k^2 \chi(k)$ curves. The
number of fitting parameters (8) was usually about half of the number of
independent data points ($N_{\rm ind} = 2\Delta k \Delta R/\pi = {}$14--17).
The accuracy of the fitting parameters was estimated from the Fisher information
matrix. To increase this accuracy, additional constraints (equal energy
corrections $\delta E_0$ for all shells) and known relations between coordination
numbers for the fluorite structure were used. FEFF software~\cite{15} was used
to calculate the dependence of the backscattering amplitude and phase, the
central-atom phase shift and the mean free path of a photoelectron on its wave
vector.

The excitation spectra and decay kinetics of CL were measured on station 3.1 of
SRS.~\cite{16}  This setup is equipped with 1-meter Seya-Namioka primary
monochromator and covers the excitation photon energy range of 5--35~eV.
Luminescence was detected by fast photomultiplier tube XP2020Q via interference
filter 204.0~nm having bandpass (FWHM) 20~nm and peak transmission 17.5\%. The
setup had no secondary monochromator for emission spectra measurements in deep
UV region. Standard time-correlated single-photon counting was used for
time-resolved measurements in a single-bunch mode operation of the SRS.

It is necessary to add that the experimental investigation of the Ba-impurity
luminescence in SrF$_2$:Ba and CaF$_2$:Ba is not a simple task because of rather
low concentration of Ba in our samples and high intensity of other emissions
spectrally overlapping with impurity CL, in particular, emissions from
self-trapped excitons of the host crystals and from uncontrolled impurities
(mainly of rare-earth ions). Due to this fact the excitation spectra of impurity
CL were measured as time-resolved excitation spectra within the short
($\sim$1~ns) time window immediately following the SR excitation pulse, thus
corresponding to detection of the fast component of luminescence. Additionally,
as CL in BaF$_2$ is temperature-independent up to 850~K (Ref.~\onlinecite{17})
and other unfavorable kinds of luminescence exhibit considerable thermal
quenching upon heating, all measurements related to impurity CL were carried
out at increased temperature of $T = 455$~K.

For luminescence measurements, BaF$_2$, SrF$_2$:Ba(1.0 mol.\%), and
CaF$_2$:Ba(1.0 mol.\%) single crystals were cleaved just prior to their
installation into the sample chamber. Before the EXAFS measurements the samples
were powdered, sieved and the powder was rubbed into the surface of adhesive
tape. From 5 to 8 layers of the tape were used for recording of EXAFS spectra.

\section{Results and discussion}

\subsection{First-principles calculations}

\begin{table}
\caption{Structural relaxation around Ba atoms in CaF$_2$:Ba and SrF$_2$:Ba solid
solutions. The experimental Ba--$X$ distances (in~{\AA}) obtained from the EXAFS
data analysis are compared with interatomic distances calculated from first
principles and interatomic distances in undoped CaF$_2$ ($a = 5.4712$~{\AA}) and
SrF$_2$ ($a = 5.8000$~{\AA}) host materials.}
\begin{tabular}{l@{~~~}l@{~~~}l@{~~~}l}
\hline
Shell & EXAFS       & First-principles & Distances in \\
      & experiment  & calculations     & host materials \\
\hline
\multicolumn{4}{c}{CaF$_2$:Ba} \\
Ba--F  & 2.568(24)   & 2.581            & 2.369 \\
Ba--Ca & 3.946(31)   & 3.942            & 3.869 \\
Ba--F  & 4.582(51)   & 4.586            & 4.536 \\
\multicolumn{4}{c}{SrF$_2$:Ba} \\
Ba--F  & 2.633(17)   & 2.628            & 2.511 \\
Ba--Sr & 4.105(91)   & 4.139            & 4.101 \\
Ba--F  & 4.779(40)   & 4.832            & 4.809 \\
\multicolumn{4}{c}{BaF$_2$} \\
Ba--F  & 2.685(14)   & ---              & 2.685 \\
Ba--Ba & 4.384(14)   & ---              & 4.384 \\
Ba--F  & 5.141(14)   & ---              & 5.141 \\
\hline
\end{tabular}
\end{table}

First-principles calculations of the local structure of SrF$_2$:Ba and CaF$_2$:Ba
solid solutions were performed within the density-functional theory in the
local density approximation (LDA) using the ABINIT software. The pseudopotentials
needed for calculations were constructed using the OPIUM program or borrowed from
Ref.~\onlinecite{18}. The local structure of the solid solutions was modeled with
CaF$_2$ and SrF$_2$ supercells containing 3$\times$3$\times$3 FCC unit cells in
which one of 27~metal atoms was substituted by the Ba atom. The cut-off energy
was 30~Ha (816~eV); the integration over the Brillouin zone was performed on the
4$\times$4$\times$4 Monkhorst-Pack mesh. Interatomic distances in the local
environment of the Ba atom, corrected for the systematic errors in the lattice
parameter determination for undoped CaF$_2$ and SrF$_2$ in the LDA, are presented
in Table~1.

As the possibility of the local phase transition into the orthorhombic phase is
also discussed in this work, we performed the first-principles calculations
for the 1$\times$2$\times$1 supercell of the $Pnma$ phase of CaF$_2$:Ba and
SrF$_2$:Ba, in which one Ca(Sr) atom was substituted by the Ba atom. The
calculations showed that the energy of this phase exceeds that of the fluorite
phase with the same Ba content by 54~meV per formula unit for CaF$_2$:Ba and by
69~meV for SrF$_2$:Ba. The atomic coordinates obtained in these calculations were
used for simulating the theoretical EXAFS spectra for the orthorhombic phase.

The energy band structure of BaF$_2$ and its solid solutions was calculated in
both the LDA and one-shot $GW$ approximations first without taking into account
the spin-orbit coupling of the Ba($5p$) states and then was corrected for the
spin-orbit splitting. The details of such calculations are described in
Ref.~\onlinecite{19}. The calculated LDA band gaps for 3$\times$3$\times$3
supercells of CaF$_2$:Ba and SrF$_2$:Ba are 7.29 and 7.03~eV, respectively; the
LDA band gap of BaF$_2$ is 6.69~eV. In all three compounds, the bottom of the
conduction band is at the $\Gamma$ point and the top of the valence band is at
the $X$ point of the Brillouin zone. In the LDA approximation, the energy
separations between the top of the Ba$^{2+}$($5p$) band and the conduction band
edge are 14.13, 14.81, and 15.22~eV in BaF$_2$, SrF$_2$:Ba, and CaF$_2$:Ba,
respectively. The calculated spin-orbit splitting of the Ba ($5p$) states in
BaF$_2$ is 2.05~eV. In order to take into account the many-body effects, the
$GW$ corrections to the positions of the valence and conduction bands in undoped
CaF$_2$ and SrF$_2$ as well as to the position of the Ba$^{2+}$($5p$) band in
BaF$_2$ were calculated. After taking into account the many-body effects and
spin-orbit splitting of the Ba($5p$) states, the final energy separations
between the top of the Ba$^{2+}$($5p$) band and the conduction band edge are
17.35, 18.23, and 18.73~eV in BaF$_2$, SrF$_2$:Ba, and CaF$_2$:Ba, respectively.

\subsection{EXAFS data}

\begin{figure}
\includegraphics{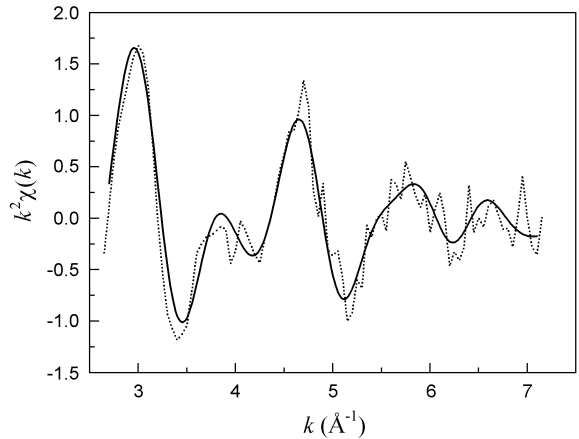}
\caption{Typical EXAFS spectrum of the CaF$_2$:Ba sample measured at the
Ba $L_{\rm III}$ edge at 300~K (dotted line) and its best theoretical fit
(solid line).}
\end{figure}

In order to determine the local environment of the Ba$^{2+}$ impurity, EXAFS
spectra were analyzed. Typical Fourier-filtered EXAFS spectrum $k^2 \chi(k)$ for
the CaF$_2$:Ba sample and its best theoretical fit are shown in Fig.~1. The best
agreement between the experimental and calculated data was obtained for a model
in which the Ba ions substitute for the Sr or Ca ions in the fluorite structure.
Interatomic distances for three nearest shells obtained from the EXAFS data
analysis, the data calculated from known lattice parameters of undoped host
crystals and the results of first-principles calculations for CaF$_2$:Ba and
SrF$_2$:Ba solid solutions are given in Table~1. It is seen that for BaF$_2$ the
interatomic distances obtained from the EXAFS data analysis agree well with
the X-ray data. For SrF$_2$:Ba and CaF$_2$:Ba solid solutions the nearest Ba--F
distances are shorter than in BaF$_2$, but are always much longer than
interatomic distances in host crystals. The distances obtained from
first-principles calculations agree with the experimental interatomic distances
in solid solutions. The coordination numbers are close to 8, 12, and 24 for
all samples, as expected for the fluorite structure.

\begin{figure}
\includegraphics{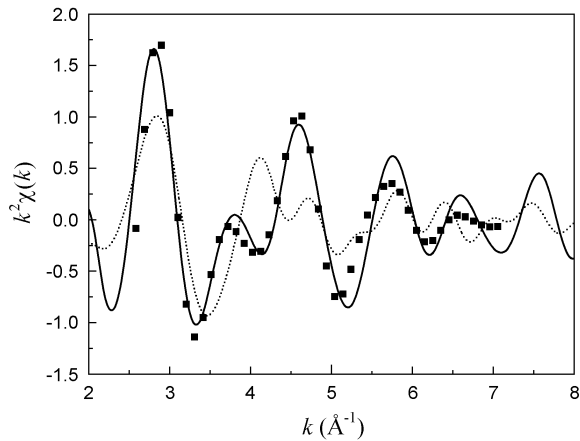}
\caption{Comparison of experimental EXAFS spectrum measured at the
Ba $L_{\rm III}$ edge for the CaF$_2$:Ba sample (points) with the spectra for
solid solutions with cubic (solid line) and orthorhombic (dotted line)
structures simulated using first-principles calculations.}
\end{figure}

On the other hand, the comparison of the experimental EXAFS spectrum with the
theoretical one simulated for the orthorhombic phase shows that the spectra are
very different (Fig.~2). So, we can conclude that the local phase transition
from cubic fluorite to orthorhombic PbCl$_2$ structure that could be induced
by the strong local stress in the host around the impurity ions is absent in
CaF$_2$:Ba and SrF$_2$:Ba samples.

As for the Debye-Waller factors, the comparison has shown that in solid
solutions these factors are close to those in BaF$_2$ (for example,
$\sigma^2_1 = 0.0144(25)$~{\AA}$^2$ in BaF$_2$ and 0.0181(81) for CaF$_2$:Ba
for the first shell). We can conclude that the possible off-centering of the
impurity Ba ion is absent in solid solutions, and so the impurity occupies the
on-center position at the site.

The obtained EXAFS results show that the impurity Ba$^{2+}$ ions substitute for
Sr$^{2+}$ and Ca$^{2+}$ ions in SrF$_2$:Ba and CaF$_2$:Ba in an on-center
position. The solid solutions do not undergo any local phase transition that
could be induced by the strong local stress around the impurity ions. On the
other hand, as the nearest Ba--F distances in SrF$_2$:Ba and CaF$_2$:Ba are
shorter compared to undoped BaF$_2$, one can expect a faster decay of CL in
SrF$_2$:Ba and CaF$_2$:Ba than in BaF$_2$ because of increased overlap of
respective wave functions.

\subsection{Luminescence spectroscopy}

The results of our EXAFS studies and first-principles calculations have shown
that the crystal structure (symmetry) around Ba$^{2+}$ ions does not change for
the case of impurity Ba$^{2+}$ ions and so, one should not expect noticeable
changes of the shape of emission spectra of the impurity CL in CaF$_2$:Ba and
SrF$_2$:Ba compared to that of intrinsic CL in BaF$_2$ (such a behavior of
emission spectra of impurity and intrinsic CL was observed, in particular, in
Ref.~\onlinecite{20} for the system
Rb$_{1-x}$Cs$_x$F). As was mentioned above, because of the low concentration of
Ba in our samples the intensity of the impurity CL from SrF$_2$:Ba and CaF$_2$:Ba
was rather low compared to that of other emissions observed from these crystals.
However, the inspection of emission spectra obtained for CaF$_2$:Ba and SrF$_2$:Ba
has shown the presence of a weak featureless band (shoulder) with a maximum at
about 220~nm at the high-energy part of the emission spectrum, which can be
associated with the impurity CL. Further investigations of excitation spectra
and decay times of this band (spectrally selected by interference filter) in the
time-resolved mode, which are described below, evidently show that the fast
luminescence observed from the studied SrF$_2$:Ba and CaF$_2$:Ba crystals is
indeed due to the impurity CL.

The CL excitation spectra measured in the region 16--32~eV for three investigated
samples are shown in Fig.~3. In BaF$_2$ the excitation of electrons from the
Ba$^{2+}$\,($5p$) core band to the conduction band starts at 18.1~eV, in agreement
with previous data.~\cite{1,2,3,4}  The thresholds of the impurity CL excitation
spectra in SrF$_2$:Ba and CaF$_2$:Ba, when electrons are excited from the
Ba\,($5p$) level of the impurity ion, are shifted to higher energy by
$\sim$0.5--1~eV. These data are in reasonable agreement with the above results
of first-principles calculations for the energy separation between the top of
the Ba$^{2+}$\,($5p$) band and the conduction band edges in BaF$_2$, SrF$_2$:Ba,
and CaF$_2$:Ba.

The case of CaF$_2$:Ba and SrF$_2$:Ba differs from that of Rb$_{1-x}$Cs$_x$F
described in Ref.~\onlinecite{20}. In the latter system the host crystal (RbF)
possesses intrinsic CL, i.e. the Rb\,($4p$) core holes do not undergo fast Auger
decay and accordingly the migration of these holes towards the nearby Cs$^+$
ions leads to the appearance of Cs\,($5p$) related impurity CL. For small
concentrations of Cs$^+$ the intensity of Cs\,($5p$) impurity CL under the
creation of Rb\,($4p$) core holes can be rather strong compared to the intensity
of Cs\,($5p$) impurity CL under direct creation of Cs\,($5p$) holes because of
the strong absorption by the host crystal. As a result, the well pronounced
excitation threshold of Cs\,($5p$) impurity CL in Rb$_{1-x}$Cs$_x$F for small
values of $x$ coincides with the edge of transitions from the host Rb\,($4p$)
band to the conduction band. In contrast to that, in SrF$_2$:Ba and CaF$_2$:Ba
the Ca\,($3p$) and Sr\,($4p$) core holes possess very fast ($10^{-14}-10^{-15}$~s)
Auger decay right after their creation which prevents the possibility of
noticeable migration of these holes towards Ba$^{2+}$ ions. Accordingly, the
excitation spectrum of impurity CL in SrF$_2$:Ba and CaF$_2$:Ba should
correspond to the spectrum of holes creation directly on the impurity
Ba\,($5p$) level, and the excitation threshold of Ba\,($5p$) impurity CL in
SrF$_2$:Ba and CaF$_2$:Ba corresponds to the edge of transitions from the
Ba\,($5p$) impurity level to the conduction band.

\begin{figure}
\includegraphics{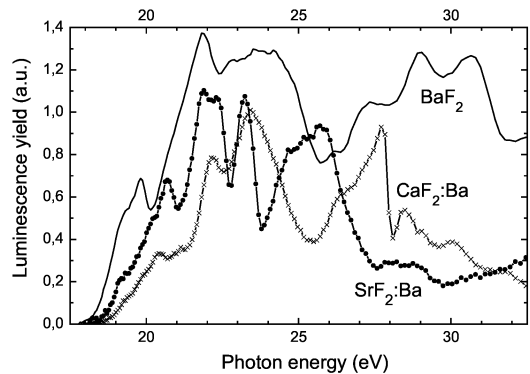}
\caption{Excitation spectra of crossluminescence for BaF$_2$ ($T = 300$~K),
SrF$_2$:Ba ($T = 455$~K), and CaF$_2$:Ba ($T = 455$~K).}
\end{figure}

Undoped BaF$_2$ (Fig.~3, curve a) shows well-known broad excitation bands
correlated to the absorption spectrum of BaF$_2$, which was extensively discussed
in previous papers.~\cite{1,2,3,4}  In SrF$_2$:Ba and CaF$_2$:Ba the impurity
Ba$^{2+}$ center should be considered as a strongly localized electronic system.
This means that for electronic transitions between the impurity Ba\,($5p$) levels
and the conduction band the momentum conservation law is not important and only
the energy conservation law should be taken into account. In this case the shape
of the absorption (excitation) spectrum will reflect the features of the density
of states in the conduction band rather than the features of band-to-band
transitions at specific points of the Brillouin zone as is the case for the
BaF$_2$ crystal. The CL excitation spectra in SrF$_2$:Ba and CaF$_2$:Ba show a
rather rich structure with several relatively narrow peaks and dips. Although
narrow dips at $\sim$22.6 and $\sim$23.7~eV in SrF$_2$:Ba and at $\sim$28.0~eV
in CaF$_2$:Ba definitely correspond to maxima in absorption (and reflection)
spectra of the host crystals and are related to the creation of the core (cation)
excitons,~\cite{21}  the structure between the excitation thresholds and the
above mentioned narrow dips corresponds to the features of absorption by impurity
Ba$^{2+}$ ions. It should be mentioned here that the reflection/absorption
spectra of SrF$_2$ and CaF$_2$ do not have any pronounced structure in the
spectral range 18--22~eV for SrF$_2$ and 18--24~eV for CaF$_2$
(Ref.~\onlinecite{20}), i.e. the host crystals are relatively transparent for
exciting radiation in these spectral ranges. Only at photon energies corresponding
to the excitation of core excitons the competition of strong absorption by the
host crystal and weaker absorption by impurity Ba$^{2+}$ ions is responsible for
well-pronounced features in the impurity CL excitation spectra.

\begin{figure}
\includegraphics{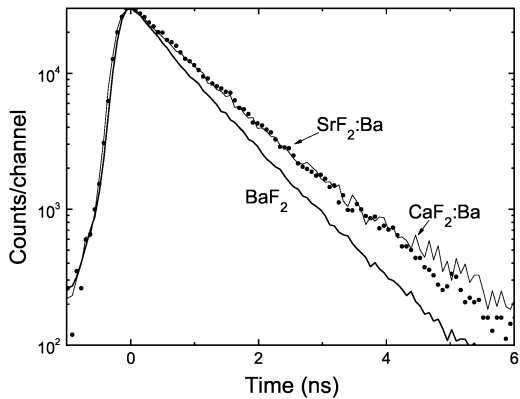}
\caption{Decay curves of crossluminescence for BaF$_2$ ($T = 300$~K), SrF$_2$:Ba
($T = 455$~K), and CaF$_2$:Ba ($T = 455$~K). Excitation energy is 30~eV.}
\end{figure}

Fig.~4 shows the CL decay curves for BaF$_2$, CaF$_2$:Ba, and SrF$_2$:Ba. In
contrast to Ba-doped crystals, the measurements for undoped BaF$_2$ were
performed at $T = 300$~K with an excitation energy $E_{\rm ex} = 30$~eV, where
the decay time of CL in BaF$_2$ has its maximum value.~\cite{22}  The fitting of
the decay curve for BaF$_2$ with a single exponent gives the value of decay time
equal to 0.79$\pm$0.01~ns. It is easily seen that CL decay time for SrF$_2$:Ba
and CaF$_2$:Ba is remarkably longer in comparison with BaF$_2$ and is equal to
0.98$\pm$0.01~ns. The measurements have revealed that CL decays in CaF$_2$:Ba
and SrF$_2$:Ba do not show any remarkable changes both with temperature (in the
temperature range 200--450~K) and excitation energy (20--32~eV). In the case of
BaF$_2$ the decay time slightly decreases with temperature above 300~K. This
fact might be explained by an increase with temperature of the band mobility of
Ba\,($5p$) core holes in BaF$_2$ resulting in an increase of the migration
losses and accordingly in acceleration of decay.

In contrast to intrinsic CL, when the uppermost core hole can migrate in the
crystal,~\cite{20,23}  one of the features of the impurity CL is that for small
concentrations ($\sim$1\%) of impurity ions the core hole created at this ion
is strongly localized on this ion and is not movable. In the case of
Ba$^{2+}$-related impurity CL in SrF$_2$ and CaF$_2$ the impurity Ba\,($5p$)
levels lie well above the uppermost core band of the host crystal (by $\sim$5~eV
in SrF$_2$ and by $\sim$10~eV in CaF$_2$ (Ref.~\onlinecite{21})). Accordingly,
thermally activated migration of the Ba\,($5p$) core hole via the host uppermost
core band is also not possible. Moreover, as was already mentioned above, the
Ca\,($3p$) and Sr\,($4p$) core holes cannot show noticeable migration because
of very fast Auger decay.

\begin{figure}
\includegraphics{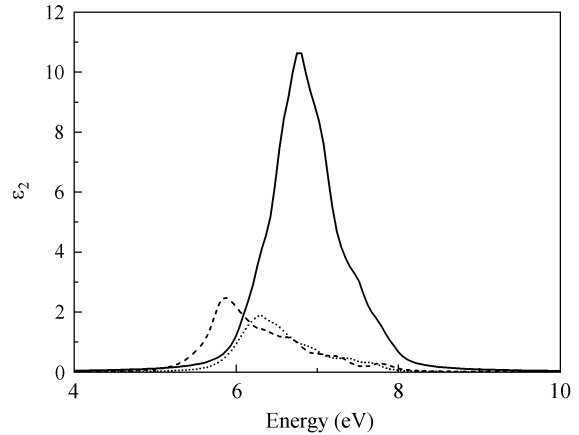}
\caption{Imaginary part of the dielectric function $\epsilon_2$ for the
$5p \to 2p$ optical transitions in BaF$_2$ (solid line), CaF$_2$:Ba (dashed
line) and SrF$_2$:Ba (dotted line). The spectra were calculated without taking
into account many-body effects and spin-orbit coupling.}
\end{figure}

In spite of the fact that the nearest interatomic Ba--F distance is shorter in
CaF$_2$:Ba and SrF$_2$:Ba in comparison with BaF$_2$ and hence the probability
of CL transitions F$^-$\,($2p$)$\to$Ba$^{2+}$\,($5p$) should be higher due to
the expected increase of the overlap of the F\,($2p$) and Ba\,($5p$) wave
functions, the decay kinetics in BaF$_2$ is remarkably faster. One explanation
for this discrepancy may be the possible mobility of the Ba\,($5p$) core holes
in BaF$_2$ (in contrast to SrF$_2$:Ba and CaF$_2$:Ba), which can lead to
non-radiative recombination of the holes at the crystal surface.

On the other hand, the overlap of wave functions of the Ba$^{2+}$\,($5p$) and
F$^-$\,($2p$) band states and accordingly the probability of optical transitions
between these states is determined not only by the Ba--F distance but mainly by
the shape of respective wave functions. In order to estimate the corresponding
probabilities of optical transitions, first-principles calculations of spectra
of the imaginary part of the dielectric function $\epsilon_2$ for transitions
from the completely filled Ba\,($5p$) band to unoccupied states of the valence
band were performed for BaF$_2$ and 2$\times$2$\times$2 supercells that model
the CaF$_2$:Ba and SrF$_2$:Ba solid solutions. Although these spectra (Fig.~5)
are absolutely hypothetical, nevertheless they allow estimating the probability
of optical transitions between the Ba\,($5p$) band and the valence band for wave
functions corresponding to the electronic states in real crystals. It is seen
that in solid solutions the averaged probability of these transitions is much
smaller than in undoped BaF$_2$. This result agrees with slower CL decay observed
in the solid solutions and so can also serve as its explanation. It should be
noted that the absorption coefficients calculated using the $\alpha(\omega) =
(2\omega/c)\{(1/2)[(\epsilon_1^2 + \epsilon_2^2)^{1/2} - \epsilon_1]\}^{1/2}$
equation from the complex dielectric function $\epsilon_1 + i\epsilon_2$ for
transitions between the Ba\,($5p$) band and the valence band are of comparable
magnitude with the absorption coefficients for allowed optical transitions
between the Ba\,($5p$) band and the conduction band and so can explain the very
fast kinetics of CL.

As was indicated above, we obtained experimentally that decay times of impurity
CL in SrF$_2$:Ba and CaF$_2$:Ba are roughly the same. This observation is also
in line with our conclusion that the CL transition probability is not directly
related to the Ba--F interatomic distance. As can be seen from Fig.~5, depending
on the photon energy, the transition probability in SrF$_2$(Ba) can be either
larger or smaller than that in CaF$_2$(Ba). We can propose that the particular
recombination processes in two our samples are such that their transition
probabilities are close, and so the closeness of the two decay times is just
a coincidence.

It is necessary to mention that the obtained results of first-principles
calculations are far evaluative because the detailed calculations should take
into account the excitonic effects and additional relaxation of the F ions
around the Ba ion that captured the core hole. According to Ref.~\onlinecite{24},
in BaF$_2$ the displacement of the nearest F ions towards the Ba ion with the
core hole can be as large as 0.24~{\AA}. Because of strong lattice distortion
associated with this relaxation, the probability of Ba ($5p$) core hole decay
can change significantly.

\section{Conclusions}

In the present work spectral and kinetic properties of extrinsic
crossluminescence (CL) in SrF$_2$:Ba and CaF$_2$:Ba are compared with those
of intrinsic CL in BaF$_2$ and are analyzed taking into account EXAFS data
obtained at the Ba $L_{\rm III}$ edge and results of first-principles
calculations. EXAFS studies have shown that the nearest Ba--F interatomic
distances in SrF$_2$:Ba and CaF$_2$:Ba are shorter than in undoped BaF$_2$,
but are always much longer than the Ca(Sr)--F distances in the host lattices.
The local structure analysis indicates that the impurity Ba ions substitute
for host Ca(Sr) ions in the on-center positions. The Ba--F distances obtained
for SrF$_2$:Ba and CaF$_2$:Ba from EXAFS data are very close to those obtained
from first-principles calculations. The solid solutions do not undergo any
local phase transition from cubic fluorite to orthorhombic PbCl$_2$ structure
that could be induced by the strong local stress in the host around the
impurity ions.

It was shown that spectral and kinetic properties of CL related to radiative
decay of Ba ($5p$) core hole depend remarkably on the local environment of
Ba$^{2+}$ ions in the host crystal. The shift of the CL excitation threshold
to higher energies correlates well with the results of first-principles
calculations for SrF$_2$:Ba and CaF$_2$:Ba solid solutions. The CL decay time
was revealed to be longer in SrF$_2$:Ba and CaF$_2$:Ba than in BaF$_2$ in
spite of expected acceleration of decay with decreasing Ba--F distance in
SrF$_2$:Ba and CaF$_2$:Ba due to the assumed increase of the overlap of wave
functions. This discrepancy was associated with the mobility of the Ba\,($5p$)
core hole in BaF$_2$, which recombines non-radiatively at some defect states
(e.g. at the surface of the crystal). The slower CL decay in SrF$_2$:Ba and
CaF$_2$:Ba solid solutions can also be associated with decreasing the
probability of optical transitions between the Ba\,($5p$) states and the
valence band predicted by first-principles calculations for solid solutions.
The latter hypothesis needs further justification by taking into account the
excitonic effects and additional relaxation of the F ions around the Ba ion
containing the core hole.

\begin{acknowledgments}
The authors would like to thank D. A. Shaw and K. C. Cheung for their
assistance when performing joint experiments at SRS. This work was partially
supported by RFBR Grant 13-02-91179.
\end{acknowledgments}


%

\end{document}